\documentclass{article}
\usepackage{graphicx}
\usepackage{authblk}

\date{}

\begin{document}

\title{Personalized Event-Based Surveillance and Alerting Support for the Assessment of Risk\footnote{International Meeting on Emerging Diseases and Surveillance. IMED 2011 \--- POSTER SESSION \--- Vienna, Austria. February 4-7, 2011.}}
\author[1]{Avar\'{e} Stewar}
\author[2]{Ricardo Lage}
\author[1]{Ernesto Diaz-Aviles}
\author[2]{ \authorcr Peter Dolog}
\affil[1]{L3S Research Center / LUH. Hannover, Germany \--- \{stewart, diaz\}@L3S.de}
\affil[2]{Aalborg University. Aalborg, Denmark \--- \{dolog, ricardol\}@cs.aau.dk}
 
\maketitle

\section*{Abstract}
In a typical Event-Based Surveillance setting, a stream of web documents is continuously monitored for disease reporting. A structured representation of the disease reporting events is extracted from the raw text, and the events are then aggregated to produce signals, which are intended to represent early warnings against potential public health threats.

To public health officials, these warnings represent an overwhelming list of  ``one-size-fits-all''  information for  risk assessment. To reduce this overload, two techniques are proposed. First, filtering signals according to the user's preferences (e.g., location, disease, symptoms, etc.) helps reduce the undesired noise. Second, re-ranking the filtered signals, according to an individual's feedback and annotation, allows a user-specific, prioritized ranking of the most relevant warnings.

We introduce an approach that takes into account this two-step process of: 1) filtering and 2) re-ranking the results of reporting signals.  For this, Collaborative Filtering and Personalization are common techniques used to support users in dealing with the large amount of information that they face.

We demonstrate the use of a multi-interest profile, which compactly allows users to define numerous criteria for filtering. Profiles can be decomposed by the system, so multiple interests can be automatically generated from the composite profile and be used to filter out undesired information.

A key result is tackling the problem of equally ranked signals, by exploiting the information within the underlying document and metadata provided by the user, such as annotations and favorite items. This metadata are exploited to learn the user's interests. Moreover, by combining multiple sources (e.g., annotations, favorite items, external Web 2.0 and multimedia data) a more comprehensive computational view of the user can be built to help re-rank results according to system-learned user preferences.

We introduce a method to allow users to explicitly build a profile to facilitate access to personalized health events, alerts and apply the profile for personalized ranking for different users or groups to support filtering of signals and associated documents. 

\begin{description}
  \item[Keywords:] biosurveillance, epidemic intelligence, personalization
\end{description}

\section*{Acknowledgments}
This work is supported by the European Community's Seventh Framework Program (FP7/2007-2013) Medical Ecosystem: Personalized Event-Based Surveillance grant number ICT 247829. http://www.meco-project.eu/.

\end{document}